\documentclass[a4paper]{jpconf}
\usepackage{graphicx}
\begin{document}
\title{TREX-DM: a low background Micromegas-based TPC for low mass WIMP detection}
\author{F.J.~Iguaz, J.G.~Garza, F.~Aznar\footnote{Present address: Centro Universitario de la Defensa,
        Universidad de Zaragoza, Zaragoza, Spain},
        J.F.~Castel, S.~Cebri\'an, T.~Dafni, J.A.~Garc\'ia, H.~G\'omez, D.~Gonz\'alez-Diaz,
        I.G.~Irastorza, A.~Lagraba, G.~Luz\'on, A.~Peir\'o and A.~Rodr\'iguez}
\address{Laboratorio de F\'isica Nuclear y Astropart\'iculas, Universidad de Zaragoza, Spain.}
\ead{iguaz@unizar.es}

\begin{abstract}
Dark Matter experiments are recently focusing their detection techniques in low-mass WIMPs, which requires the use
of light elements and low energy threshold. In this context, we present the TREX-DM experiment,
a low background Micromegas-based TPC for low-mass WIMP detection.
Its main goal is the operation of an active detection mass $\sim$0.300 kg, with an energy threshold below 0.4 keVee
and fully built with previously selected radiopure materials. This article describes the actual setup,
the first results of the comissioning in Ar+2\%iC$_4$H$_{10}$ at 1.2 bar and the future updates for a possible physics run
at the Canfranc Underground Laboratory in 2016.
A first background model is also presented, based on Geant4 simulations and a muon/electron discrimination method.
In a conservative scenario, TREX-DM could be sensitive to DAMA/LIBRA and other hints of positive WIMPs signals,
with some space for improvement with a neutron/electron discrimination method or the use of other light gases.
\end{abstract}

\section{Motivation}
The detection of Dark Matter (DM) \cite{Baudis:2012lb} is one of the open challenges of Astroparticle
and Particle Physics for the next years.
Evidence for DM is well founded in different observations like the anisotropies in the cosmic microwave background,
the distribution of matter in our galaxy or its gravitational effect on visible matter. However, its nature is still unknown
but its solution may involve new particles with masses and cross-sections characteristic of the electroweak scale.
The most compelling ones are the weakly interactive massive particles (WIMPs), which is a generic class of DM
candidates \cite{Lee:1977bl}, or axions and axion-like particles (ALPs) \cite{Baker:2013kb}. From the first type,
some positive hints of signals have been reported.
The most important one is due to DAMA/LIBRA experiment \cite{Bernabei:2013rb},
which has observed an annual modulation (14 cycles) compatible with that expected for galactic halo WIMPs.
More recently, CoGeNT \cite{Aalseth:2011cea} and CDMS-II \cite{Agnese:2013ra2}
have announced other possible positive signals compatible with a WIMP halo but other analysis \cite{Davis:2014jhd, Agnese:2014ra}
have respectively attributed them to a detector effect or an incomplete background model.
These hints are in contradiction with the results of other experiments like LUX \cite{Akerib:2014dsa},
supossing an spin-independent isosping-invariant WIMP-nucleon coupling
and conventional astrophysical assumptions for the WIMP halo.

\medskip
The current strategy of DM experiments is based on accumulating large target masses
of heavy nuclei (like Xenon), keeping low background levels by a systematic radiopurity control of all components
and an enhancement of the electron/neutron discrimination methods.
However, this strategy is not suitable for low WIMPs masses as for heavy nuclei this translates into very
low nuclear recoil energies, and discrimination methods normally fix a relatively high energy threshold (1-10 keV).
Given that low WIMP masses are invoked to explain the reported hints, there is now strong interest in
exploring new strategies better focused on the low mass range.
For instance, CDMSlite experiment \cite{Agnese:2013ra} has recently reached an energy threshold as low as 90 eVnr
in Germanium detectors.

\medskip
Gaseous detectors may also play an important role in future experiments,
as they can reach sub-keV energy threshold ($\sim$ 100 eV) and have access to richer topological information.
Most of these experiments (like DRIFT \cite{Daw:2012ed} and MIMAC \cite{Santos:2012ds}) are focused on directional
detection of DM \cite{Ahlen:2013sa}, aiming to exploit the relative inhomogeneity of the WIMP incoming momenta,
with a maximum expected in the direction of the Cygnus constellation.
This stream could be observed by a low-pressure gaseous detector, as it will effectively select nuclear recoils from electrons
\cite{Billard:2012jb} and it could determine their direction and sense of the nuclear recoil tracks \cite{Billard:2012jb2}.
However, these last features fastly degrade for long drift distances,
which may limit the scalability of this type of experiments.

\medskip
In this context, the TREX-DM experiment proposes another strategy based on high gas pressures,
even if neutron/electron discrimination could be less effective, but keeping a low energy threshold.
TREX-DM is a low background Micromegas-based TPC for low-mass WIMP detection and will profit all developments
made in Micromegas technology \cite{Giomataris:1995fq, Giomataris:2006yg, Andriamonje:2010sa},
as well as in the selection of radiopure materials \cite{Cebrian:2011sc, Aznar::2013fa},
specially in CAST \cite{Aune:2014sa} and NEXT-MM \cite{Alvarez:2014va} projects.
Its main goal is the operation of an active detection mass $\sim$0.300 kg
with an energy threshold below 0.4~keVee (as already observed in \cite{Aune:2014sa}) or lower.
Another experiment with a similar approach is SEDINE \cite{Gerbier:2014gg},
a Spherical Proportional Counter filled with a neon-helium mixture at high pressure
and that has reported an energy threshold as low as 0.1~keVee. 

\medskip
This article describes the actual setup of TREX-DM and the first results of the comissioning during 2014.
In a second part, a first background model of the experiment is presented, as well as its sensitivity
to low-mass WIMPs. We conclude with a summary and some prospects.

\section{Description and comissioning}
\begin{figure}[htb!]
\centering
\includegraphics[width=92mm]{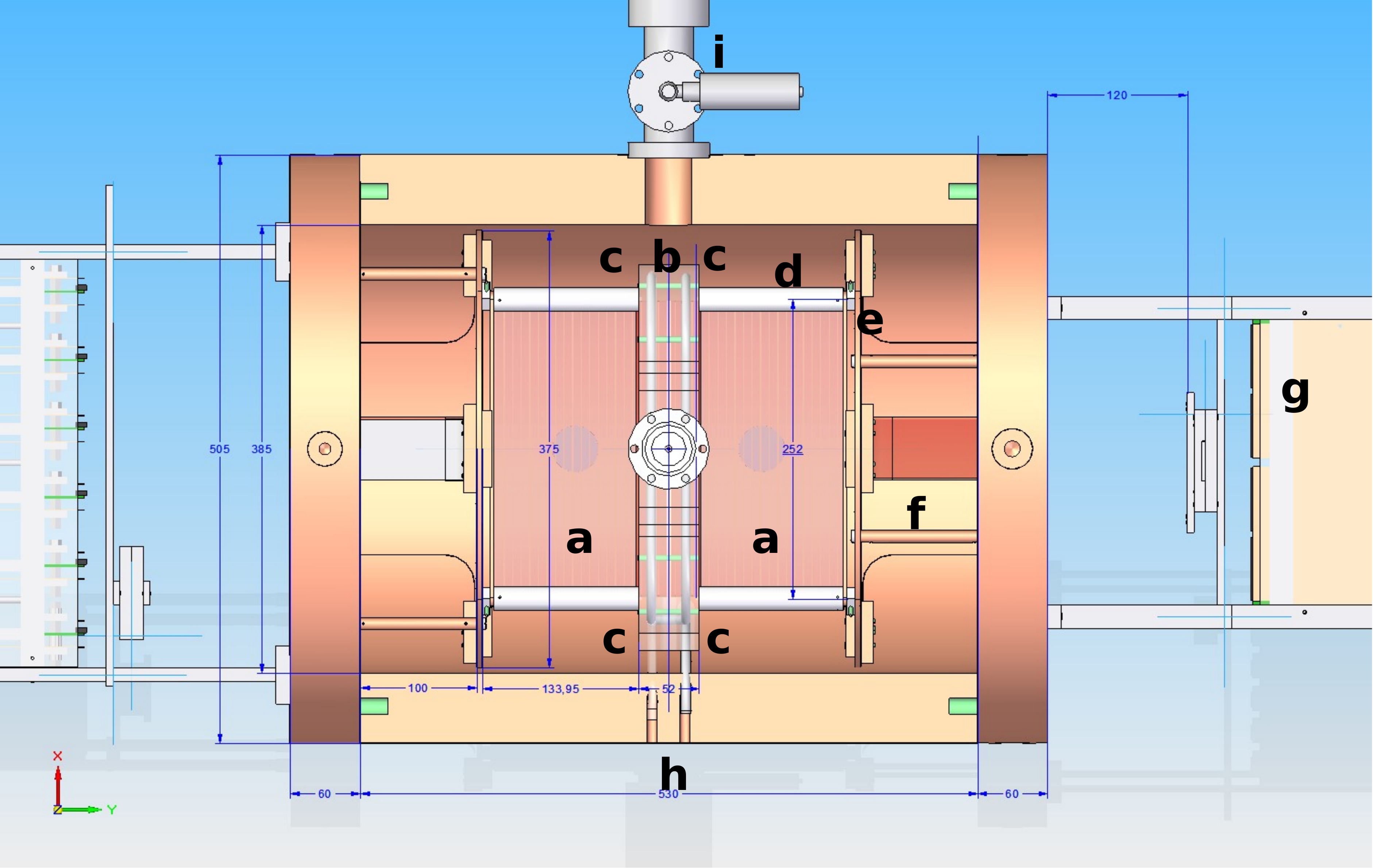}
\includegraphics[width=58mm]{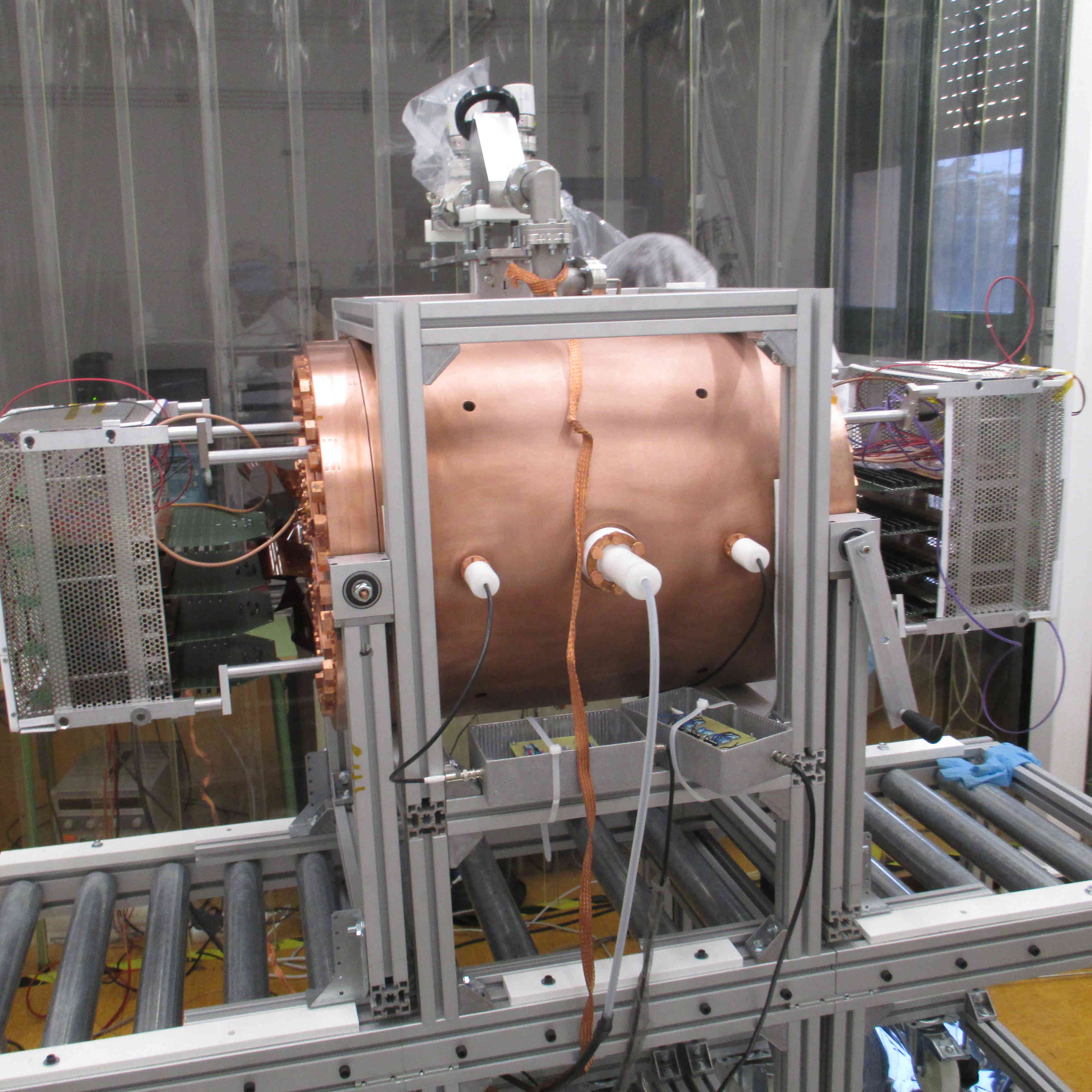}
\caption{Left: Design of the TREX-DM detector. Its different parts are described in detail in the text:
active volumes (a), central cathode (b), calibration points (c), field cage (d), Micromegas detector and support base (e),
flat cables (f), AFTER-based electronics (g), gas system (h) and pumping system (i). Right: A view of the experiment
during the comissioning.}
\label{fig:Setup}
\end{figure}

The actual setup (see figure \ref{fig:Setup}) is composed of a high purity copper vessel, with an inner diameter of 0.5~m
and a length of 0.5~m. The vessel's thickness is 6~cm,
so that it could shield most external radiation if combined with an external 10~cm thick lead layer.
The vessel contains two active volumes ({\it a} in the design), separated by a central copper cathode ({\it b}). At each side
there is a field cage ({\it d}) that makes uniform the drift field along the 19~cm between the cathode and the detector.
Each Micromegas detector ({\it e}) is screwed to a copper base, which is then attached to the vessel's inner walls by means of
four columns. The gas enters the vessel by a feedthrough at the bottom part ({\it h}) and comes out by another one
at the top part ({\it i}). The calibration system consists of a plastic tube entering in the bottom part ({\it h}), which
allows to calibrate each side at four different points ({\it c}) with a $^{109}$Cd source,
emitting X-rays of 22.1 (K$_\alpha$) and 24.9 keV (K$_\beta$).
Finally, the vessel can be pumped from the feedthrough ({\it i}) at the top to reduce the outgassing rate from the inner walls.

\medskip
The design of the two Micromegas detectors is a modified version of the CAST-MM one.
Each detector is on a Printed Circuit Board (PCB) of 1.6~mm thickness. The active surface of $25 \times 25$ cm$^2$
is divided in squared pads of 400~$\mu$m length and separated by a pitch of 500~$\mu$m.
Pads are alternatively interconnected in X and Y directions (400~strips per direction) through metallized holes,
which are rooted into four connectors prints at the sides of the PCB.
Instead of using a leak-tight PCB as in MIMAC detector \cite{Iguaz:2011fa}, a flat cable is linked to
each connector footprint by means of a SAMTEC connector (GFZ 300 points). The connectivity is assured by four screws,
which also join two 1~cm thick lead pieces and two 1~cm-thick copper containers to shield the contamination of
SAMTEC connectors \cite{Aznar::2013fa}. Each flat cable goes out from the vessel by means of a copper feedthrough.

\begin{figure}[htb!]
\centering
\includegraphics[width=50mm]{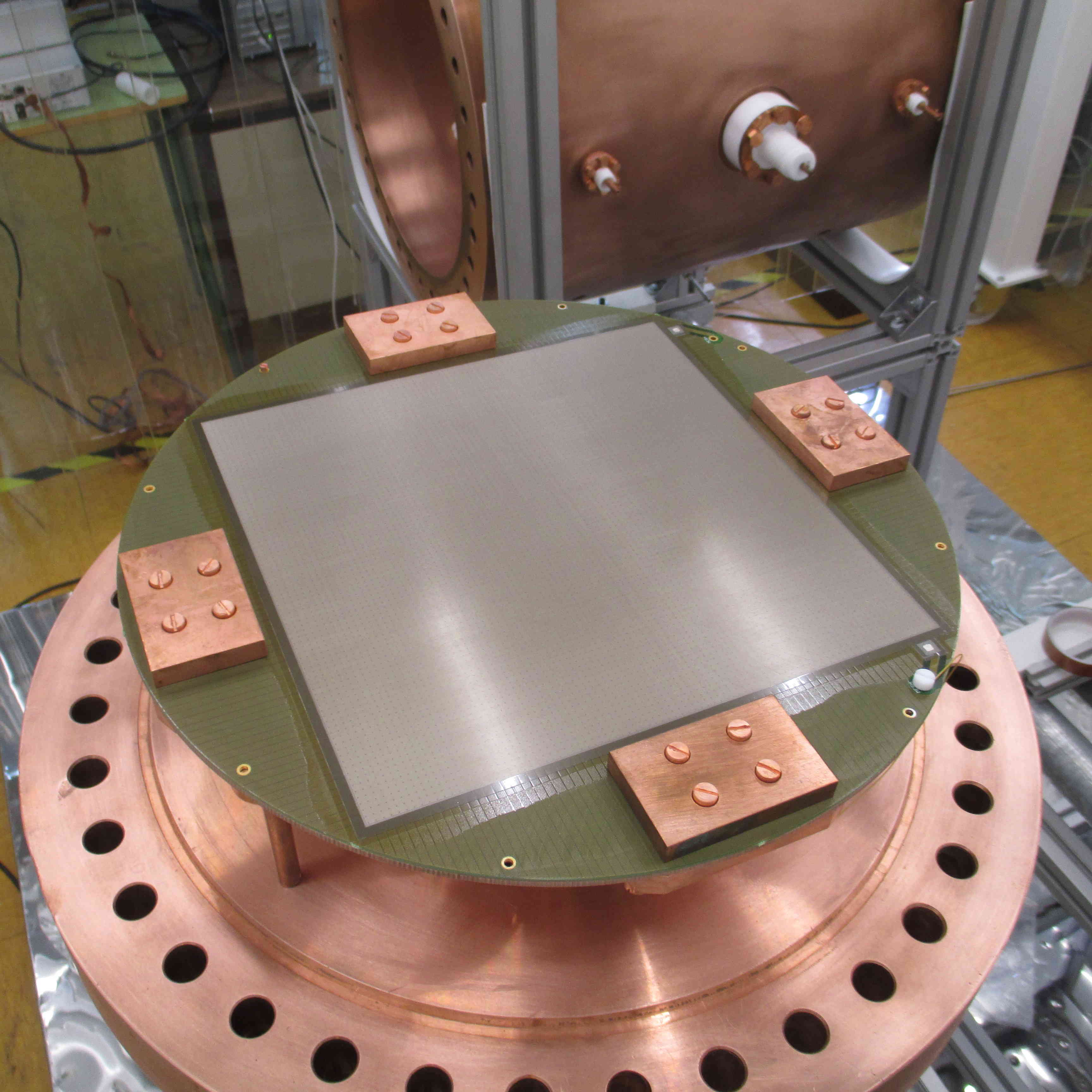}
\includegraphics[width=50mm]{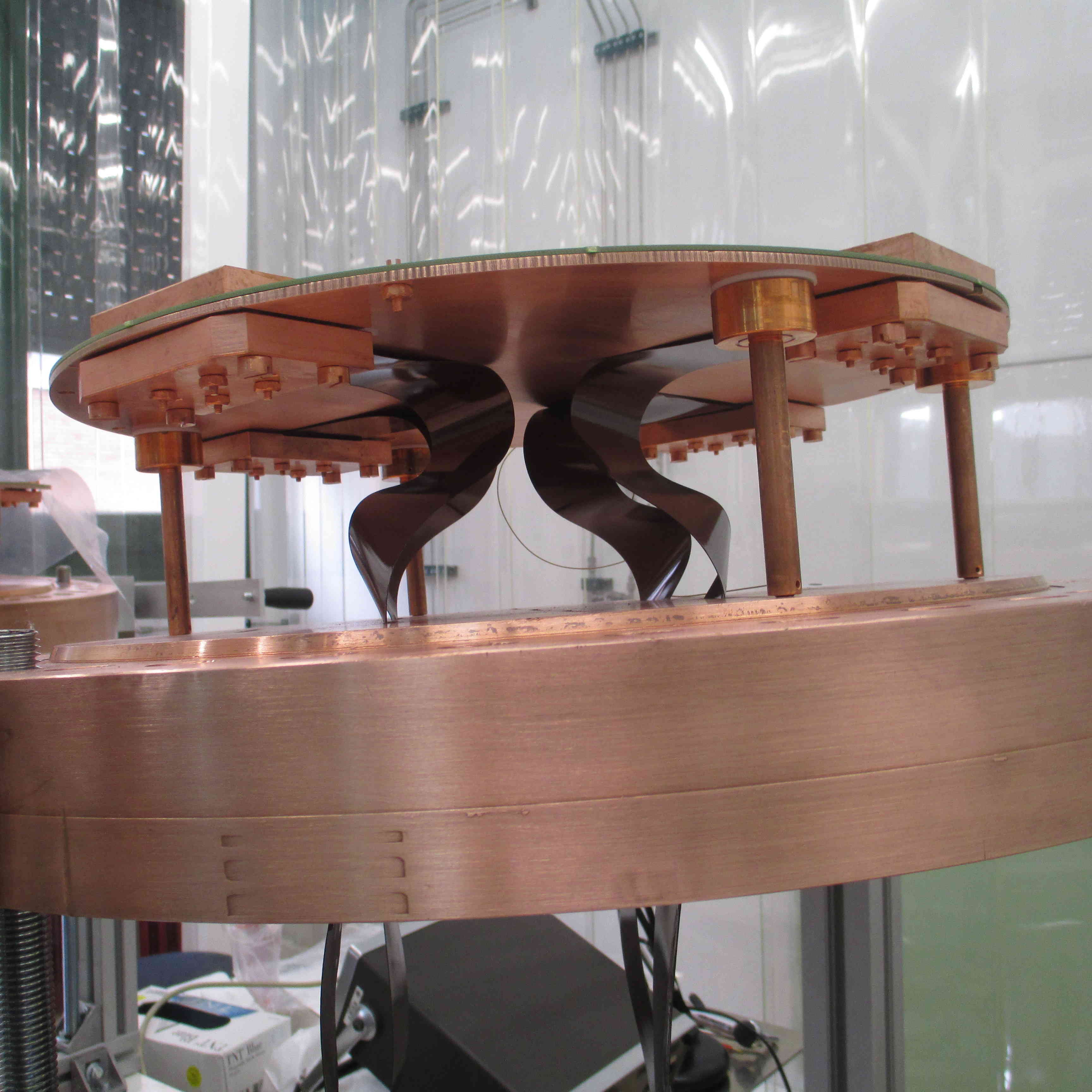}
\includegraphics[width=50mm]{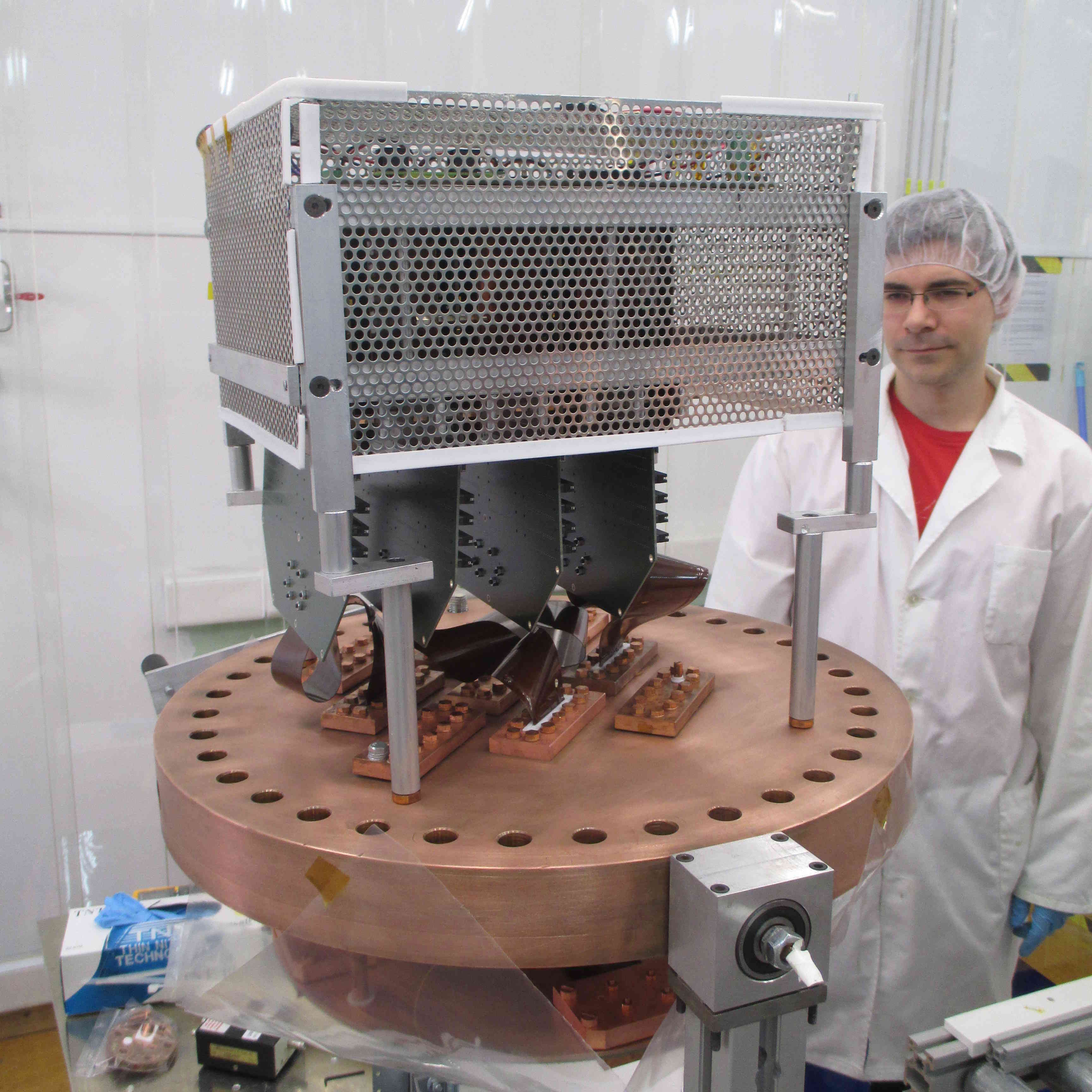}
\caption{Three views of TREX-DM Micromegas detectors. Left: a bulk detector installed at its support base
with its four flat cables already linked to it by four Samtec connectors. Center: Flat cables come out from the vessel
by their corresponding feedthroughs. Right: Flat cables are linked to their interface cards, which are then
connected to the FECs boards.}
\label{fig:BulkDetector}
\end{figure}

\begin{figure}[htb!]
\centering
\includegraphics[width=50mm]{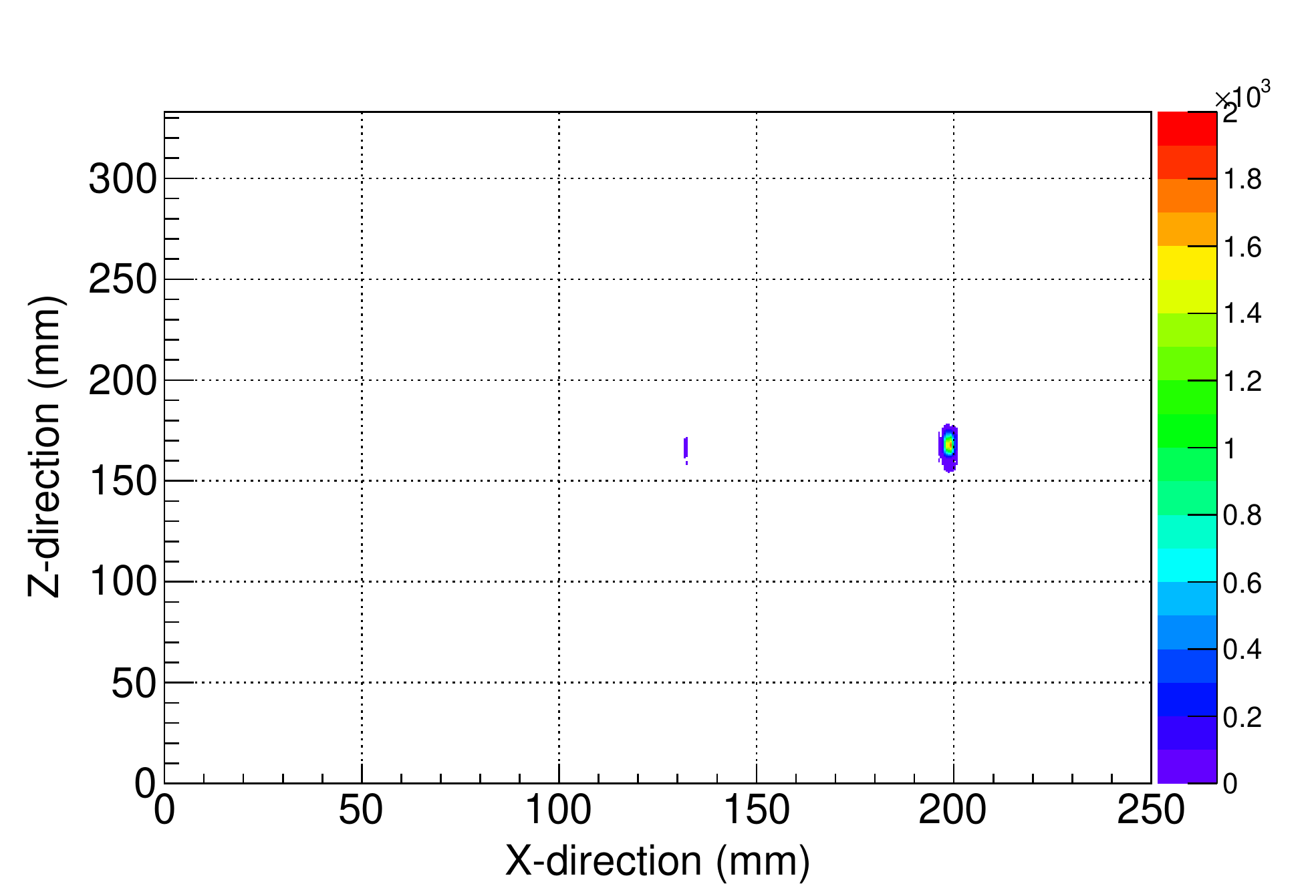}
\includegraphics[width=50mm]{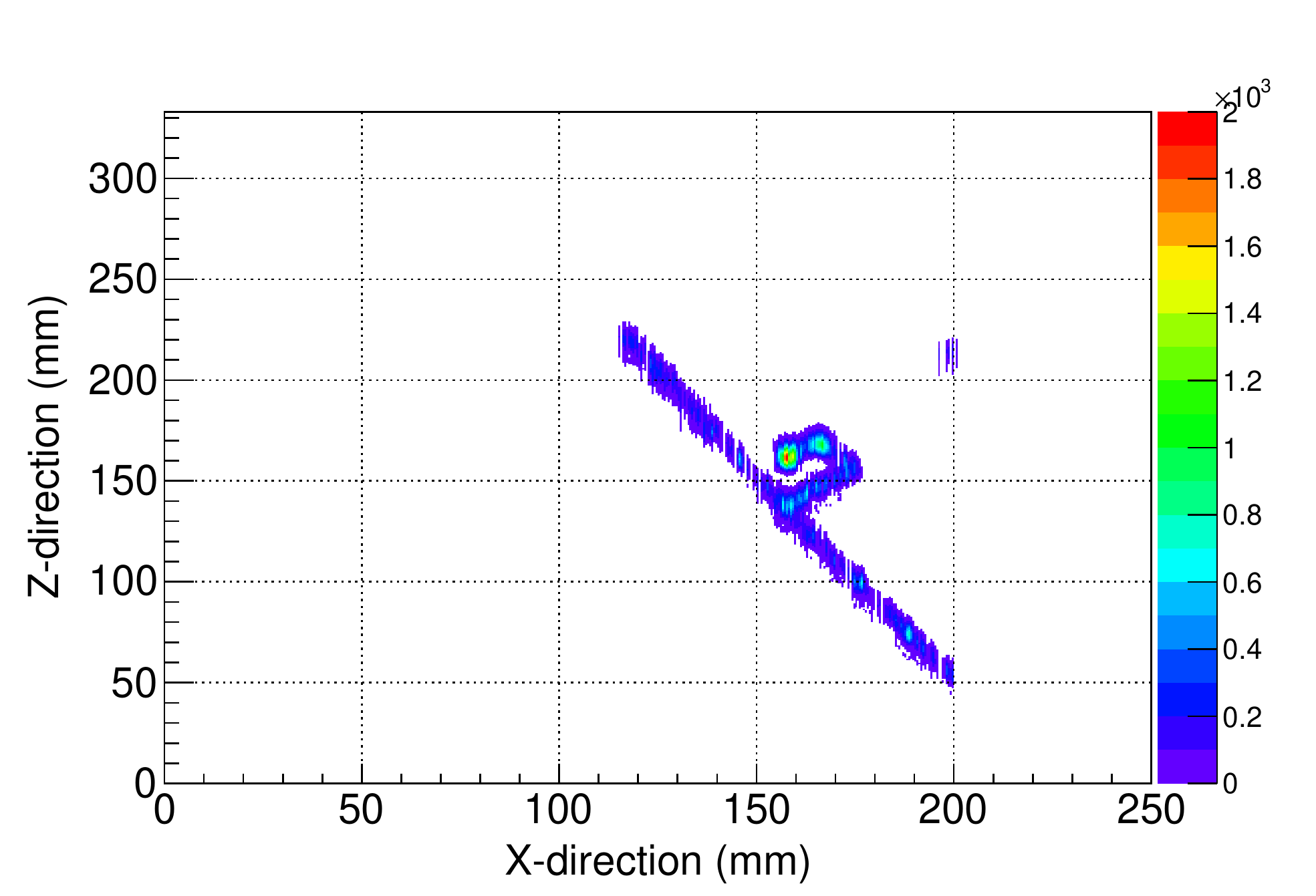}
\includegraphics[width=50mm]{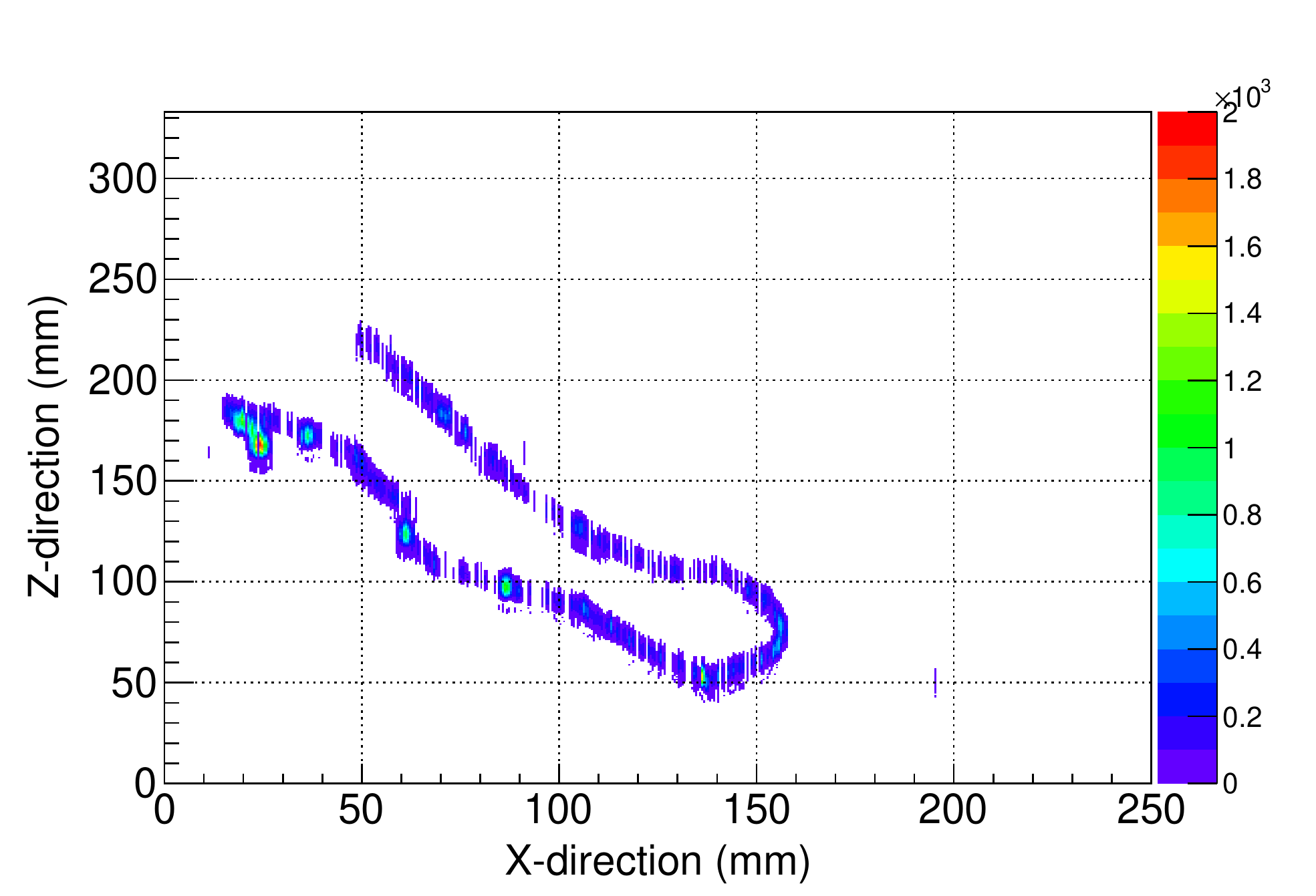}
\includegraphics[width=50mm]{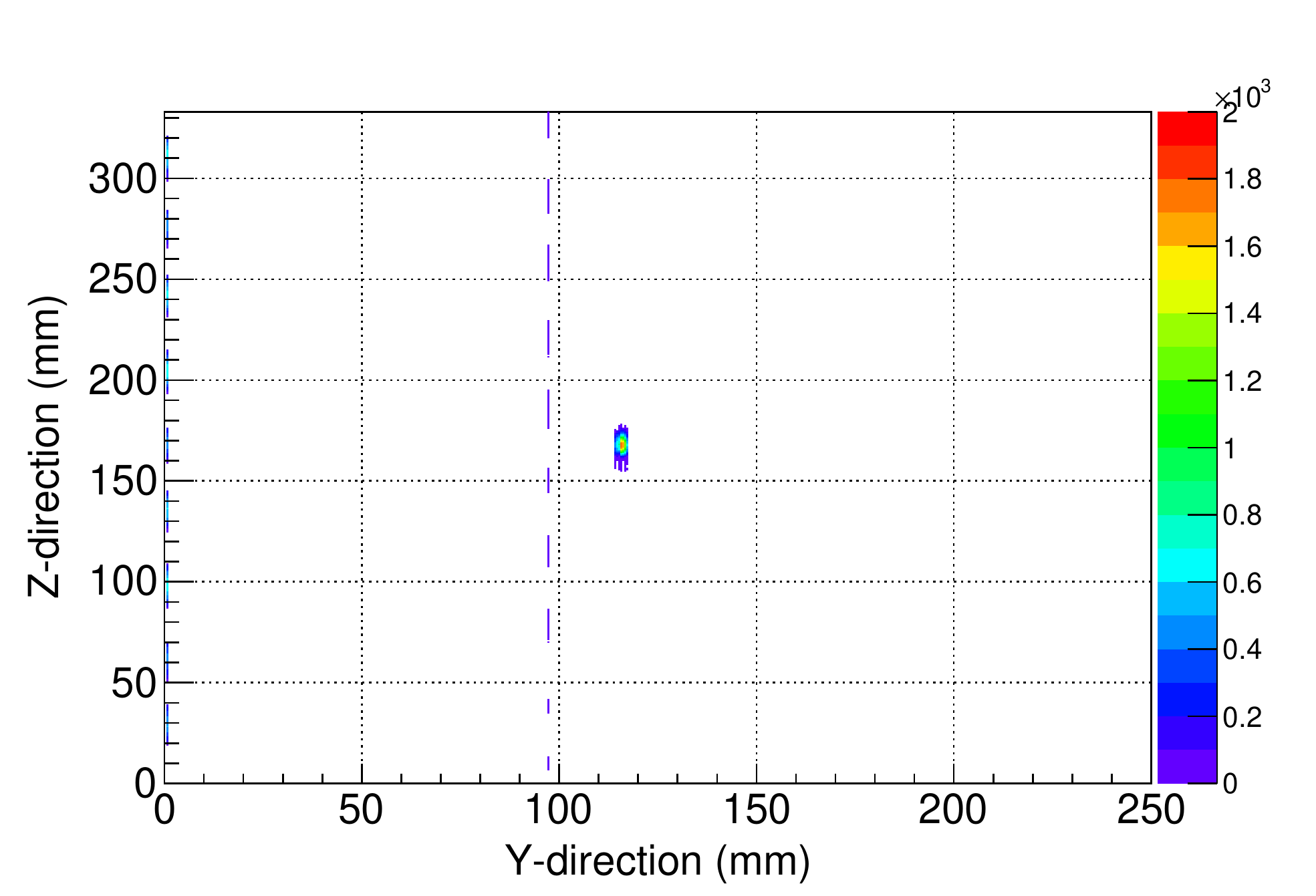}
\includegraphics[width=50mm]{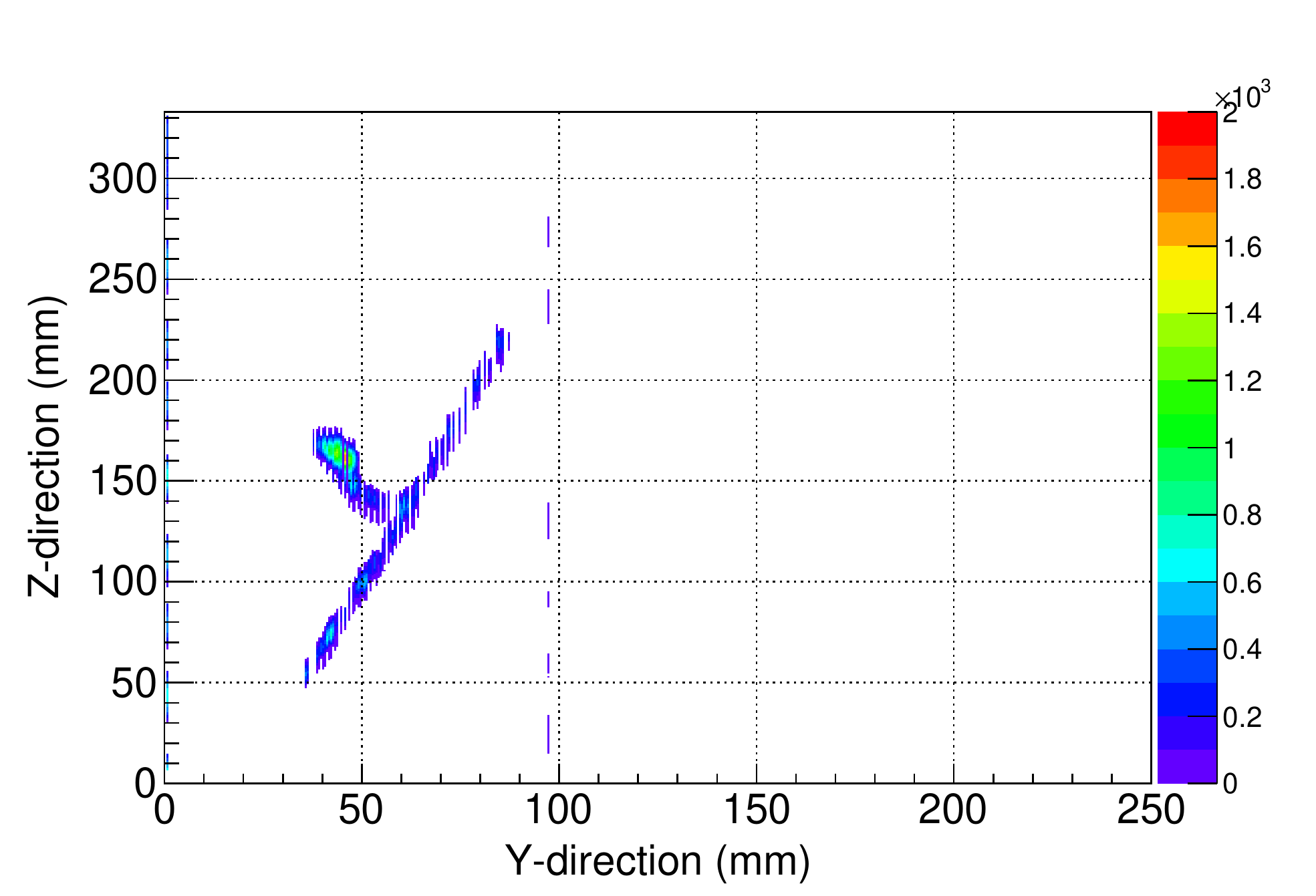}
\includegraphics[width=50mm]{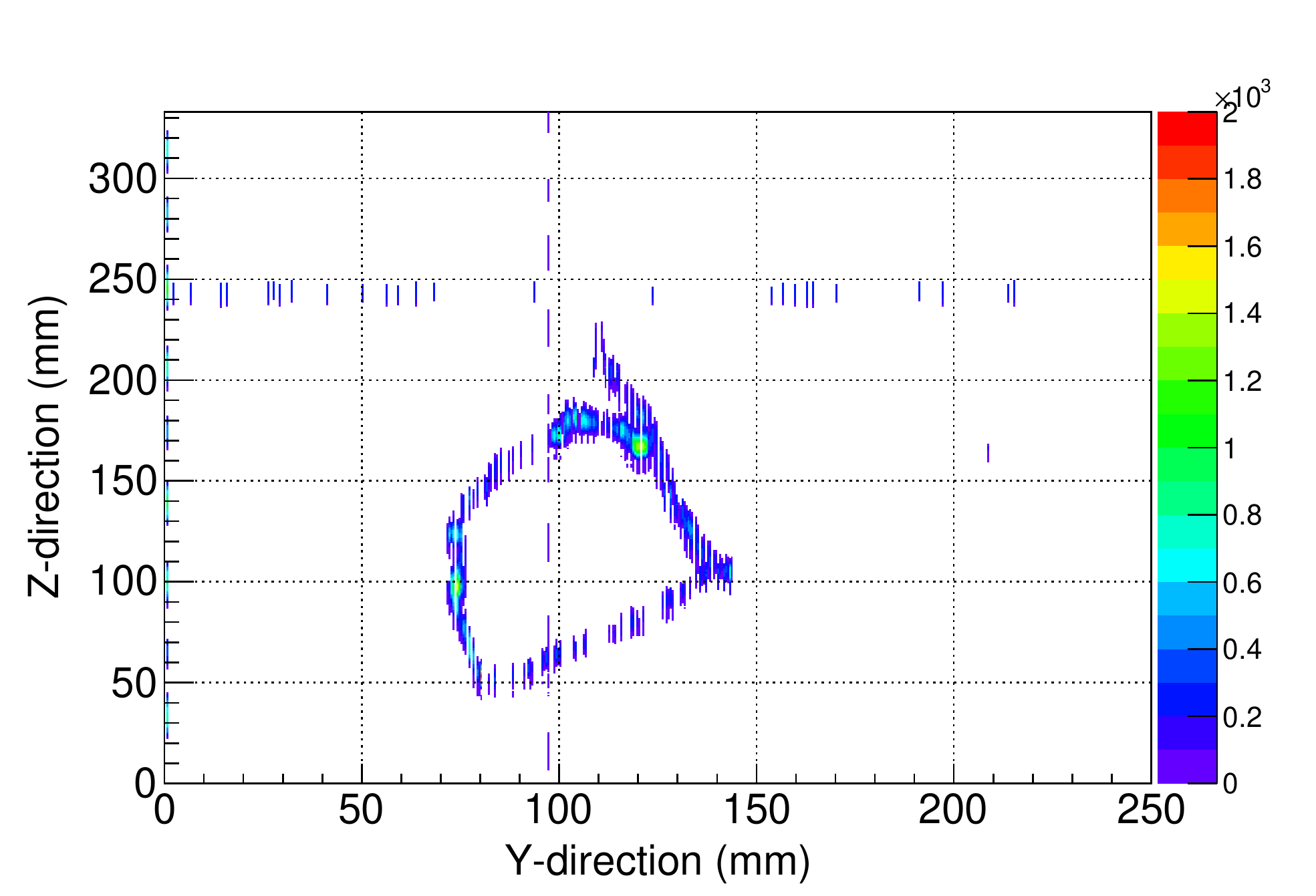}
\caption{The XZ (top) and YZ views (bottom) of three different events detected during the comissioning of TREX-DM experiment:
a low energy X-ray (left), a muon crossing the active volume which generates a $\delta$-ray (center)
and an electron with a long twisted track and a final big energy deposition or blob (right).
The color scale is proportional to the deposited charge and all x, y and z-axis are in mm.}
\label{fig:2DEvents}
\end{figure}

\medskip
The operation principle is the same as in any Micromegas-based TPC. An event interacts in the active volume and releases
some electrons, which drift toward the Micromegas plane. Electrons are then amplified in a gap of 128~$\mu$m and the
charge movement induces signals both at the mesh and the strips. The mesh signal is
extracted from the vessel by a low-voltage feedthrough and then decoupled from the high voltage
(powered by a CAEN N470A) by a capacitance.
It is then consecutively amplified by two CANBERRA preamplifier and amplifier
and later recorded by a Tektronix DS5054B oscilloscope.
In parallel, strips pulses come out from the vessel by four flat cables. Each of them is linked to an interface card
which distributes signals to the entrance connectors of
an AFTER (ASIC For TPC Electronics Readout)-based FEC board \cite{Baron:2008pb, Baron:2010pb}.
Each board has four AFTER ASICs that collect and sample the strip signals continuously at 50~MHz in 512 samples per channel,
recording a window of $\sim$10 $\mu$s, which is longer than the maximum drift time of charges created in the active volume.
The readout electronics is triggered by the negative component of the mesh's amplified bipolar pulse.
At that moment, the analog data from all channels is digitized by an ADC converter.
Finally, a pure digital electronics card, the FEM board, gathers the ADC data,
performs the pedestal substraction and sends it to the DAQ system by means of a standard network.
The XZ and YZ views of an event are reconstructed combining the strips pulses, whose temporal position gives the relative
z position, and the decoding of both the detector and the interface card.
Some examples of reconstructed events are shown in figure \ref{fig:2DEvents}.

\medskip
The TREX-DM prototype is part of the wider scope ERC-funded project called TREX \cite{TREX}, that since 2009 is devoted to
R\&D on low background TPCs and their potential applications in axion, double beeta decay and dark matter experiments.
Work on the TREX-DM prototype started 2012 with the first designs and it is now being commissioned at the TREX lab at Zaragoza.
Most of the components have been validated: the leak-tightness of all feedthroughs has been verified as their leak rate
is below $10^{-5}$ mbar l/s; the vessel is leak-tight and can keep pressures up to 10 bar; the drift cage
has been tested at high voltage in argon-based mixtures up to 10 bar; the gas flow and pressure, the temperature
and the high voltage suppliers are continuously monitored by a slow control programmed in Python
and based on Arduino cards \cite{Peiro:2013ap};
and the first signals were observed at the end of 2014 when the TPC was operated in Ar+2\%iC$_4$H$_{10}$ at 1.2 bar.
As described in last section, some components, like the detector or the electronics, will be replaced during 2015
and are only used to validate the experimental design.

\begin{figure}[htb!]
\centering
\includegraphics[width=80mm]{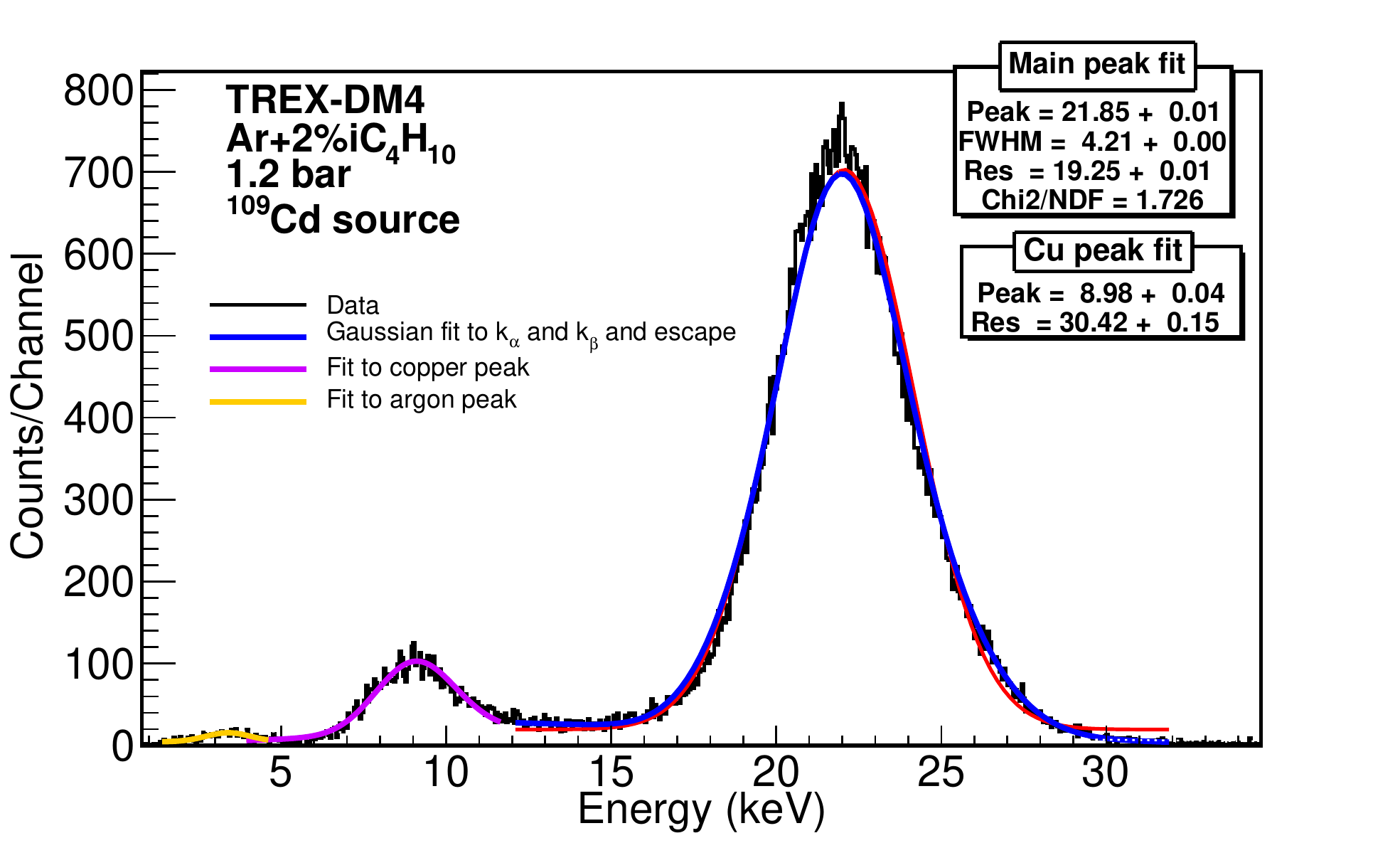}
\includegraphics[width=70mm]{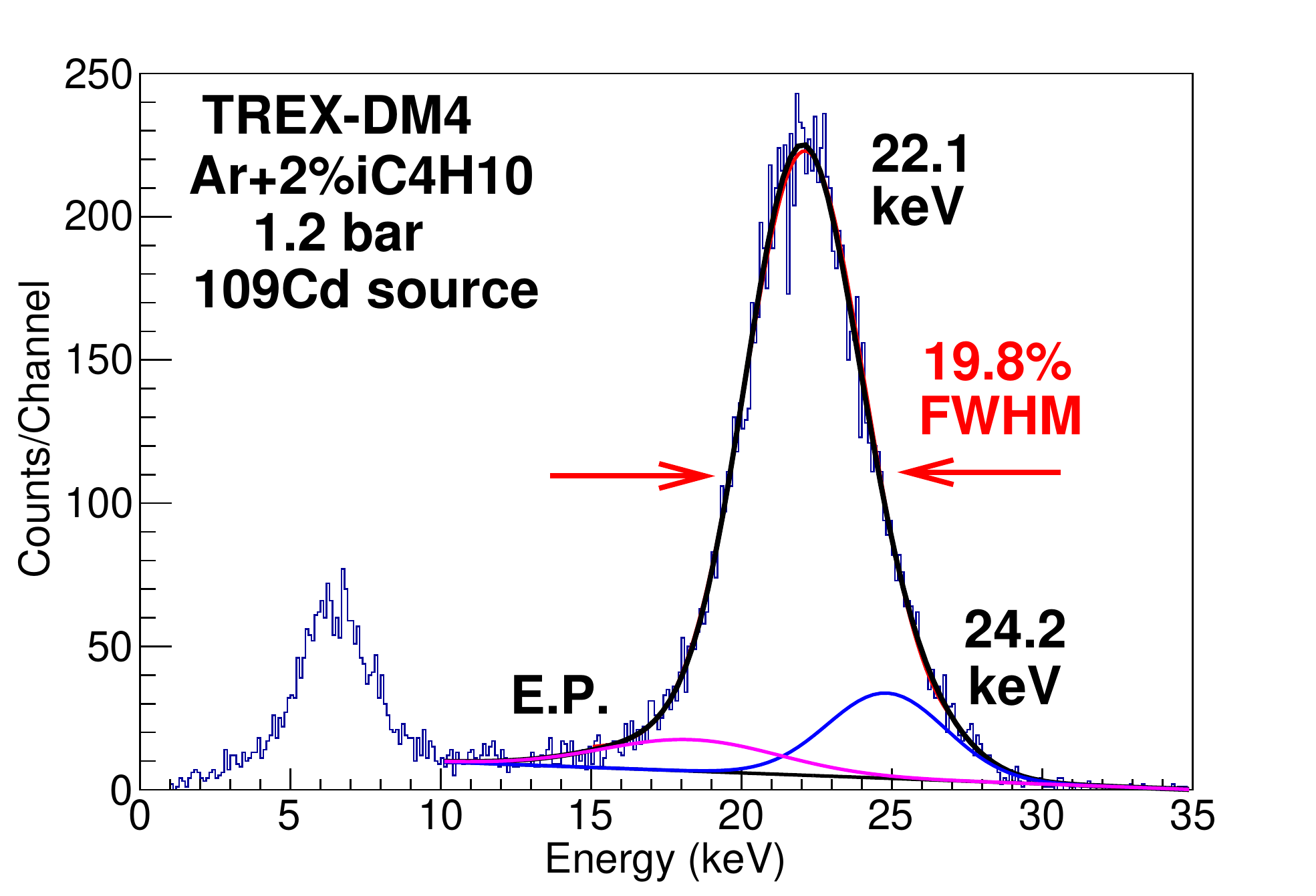}
\caption{Energy spectrum respectivelly generated by mesh (left) and strips signals (right) when one of the Micromegas
detectors was irradiated by a $^{109}$Cd source in Ar+2\%iC$_4$H$_{10}$ at 1.2 bar.
The energy resolution at 22.1 keV of both spectra has been calculated fitting the spectrum to four gaussians,
corresponding to the K$\alpha$ and K$_\beta$ lines of the source and their escape peaks (E.P.).
The fluorescence lines of iron (at 6.4 keV, emitted from the mesh) and copper (8 keV, from the vessel)
are also present in both spectra. The energy threshold is situated at 1 keV.}
\label{fig:Results}
\end{figure}

\medskip
The general performance of Micromegas detectors was studied in these conditions.
They showed a large plateau at the electron transmission curve, compatible with other bulk detectors.
However, the maximum gain reached before the spark limit ($10^3$) was far from values reached by bulk detectors ($10^4$)
and the energy resolution at 22.1~keV ($\sim$20\%~FWHM, see figure \ref{fig:Results}) was modest.
The low gain may be explained by the low quantity of quencher in gas \cite{Iguaz:2012fi}
and might be improved at higher pressures if Ar+2\%iC$_4$H$_{10}$ is then the optimum mixture \cite{Cebrian:2013sc}.
Energy resolution is clearly degraded by noise conditions during the data-taking, due to the presence of a 1~MHz
frequency. This noise was present both at mesh and strips and may be produced by a bad grounding at some point
of the electronics noise. Nevertheless, the energy threshold was situated at 1~keV.

\section{A first background model of TREX-DM}
The sensitivity of TREX-DM has been studied creating a first background model of the experiment,
if it were installed in the Canfranc Undeground Laboratory (LSC).
We have studied two light gas mixtures at 10 bar: Ar+2\%iC$_4$H$_{10}$ and Ne+2\%iC$_4$H$_{10}$;
with an active mass of 0.300 and 0.160~kg respectively.
These gases are good candidates to detect WIMPs of masses below 20 GeV. However, one of argon's isotopes ($^{39}$Ar)
is radioactive ($\beta$-decay, Q = 565~keV) with a long half-life (239~yr) and may limit the sensitivity of any
argon-based experiment. This isotope is not present in argon if it is extracted from undeground sources,
as it only appears by cosmogenic activation \cite{Acosta:2008da}. We have also simulated the main radioactive isotopes
of all the components considering their measured activities \cite{Cebrian:2011sc, Aznar::2013fa}
and the cosmic muon flux in Canfranc, at a depth of 2450 m.w.e. In the case of the Micromegas detectors and Samtec
connectors, we have considered the activities of their radiopure alternative.
The external gamma flux has not been considered as its contribution may be supressed by an external shielding.

\begin{figure}[htb!]
\centering
\includegraphics[width=60mm]{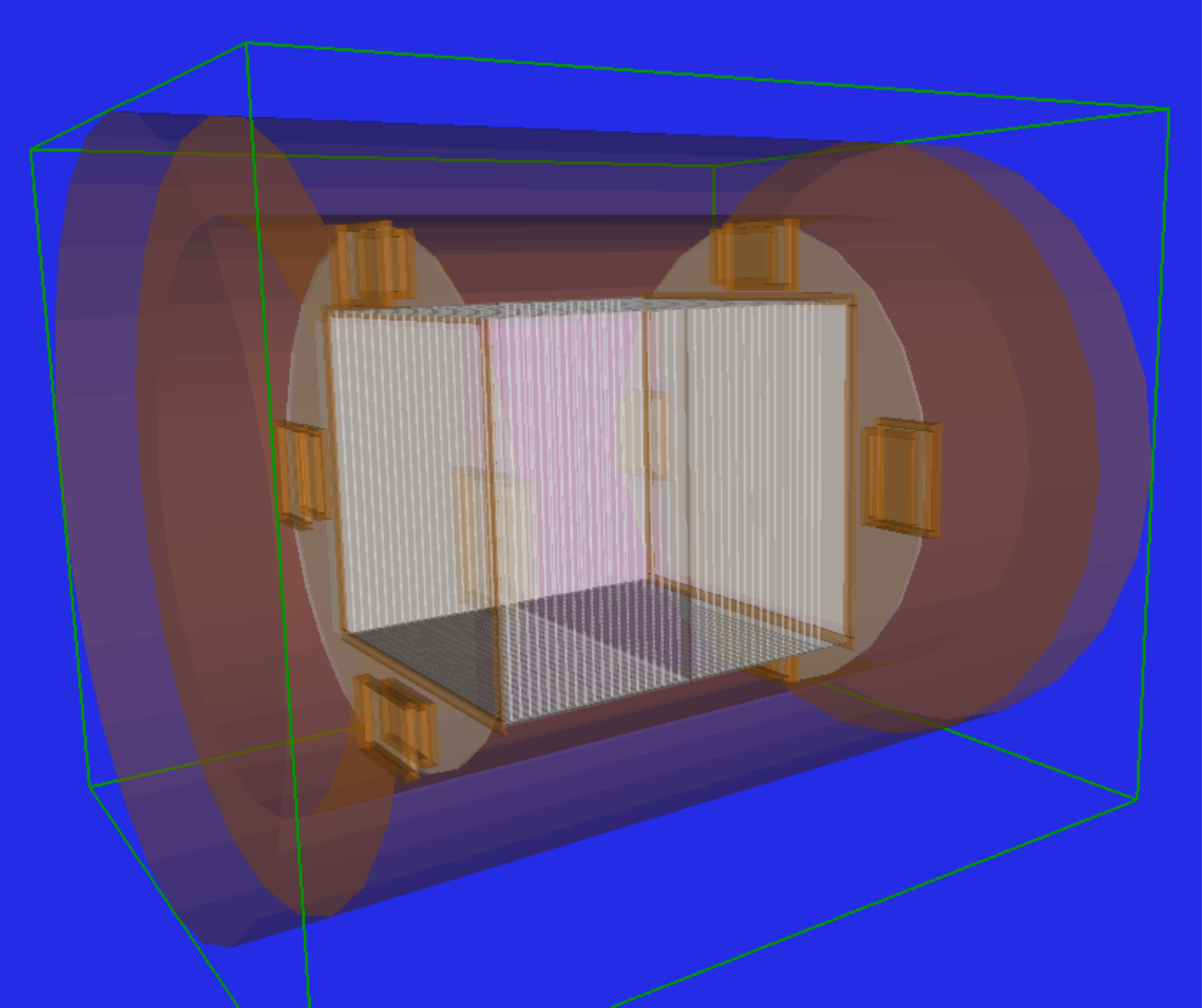}
\includegraphics[width=80mm]{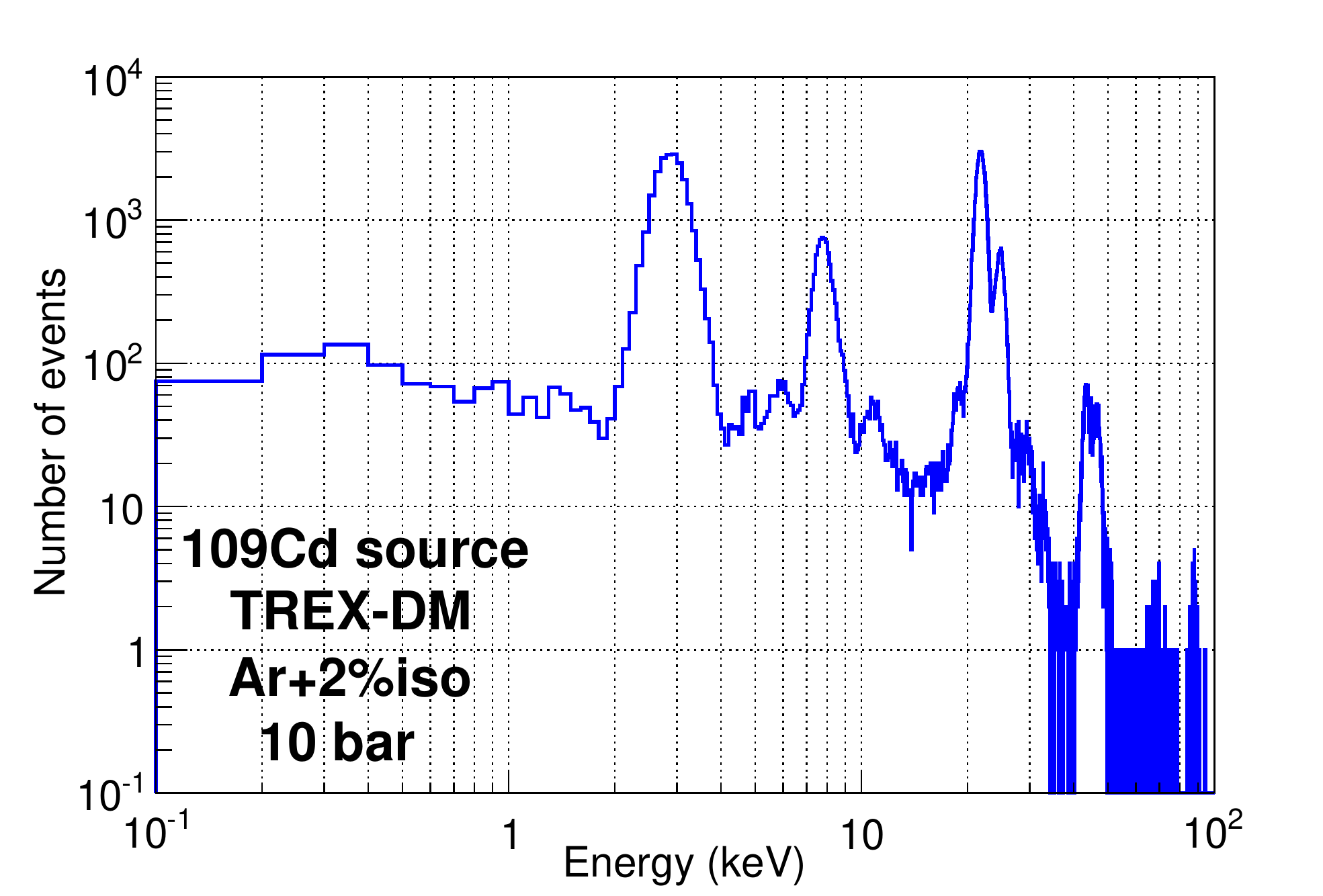}
\caption{Left: A view of the TREX-DM geometry implemented in Geant4. The cylindrical copper vessel (orange volumes)
contains a circular base with four shielded boxes (dark grey surface with four yellow boxes), 
two active volumes (in light grey), the field cage and degrador (white walls) and a central cathode.
Right: Expected energy spectrum in Ar+2\%iC$_4$H$_{10}$ at 10 bar when illuminating one side by
a $^{109}$Cd source situated at a calibration point.}
\label{fig:Geant4}
\end{figure}

\medskip
The simulation of the high pressure TPC can be divided into two blocks. The first one covers all the physical processes
involved in the passage of gamma-rays and charged particles through matter, and is based on the Decay0 \cite{Pokratenko:2000op}
and the Geant4 simulation packages \cite{Agostinelli:2003sa}. In this last program, the actual geometry has been implemented
as shown in figure \ref{fig:Geant4} (left). The second block describes the generation of electrons in the gas,
the diffusion effects during the drift to the readout plane, the charge amplification in the Micromegas gap
and the generation of signals both at mesh and strips. A detailed description can be found in \cite{Tomas:2013at}.
Some minor changes in data flow have been made to implement a two-volumes geometry, to accelerate the simulation
of diffusion process and to implement the AFTER-based electronics. The resulting data has the same format as DAQ data,
so as both real and simulated data may be analized by the same routines.
As an example, we show in figure \ref{fig:Geant4} (right) the expected energy spectrum when illuminating one side
by a $^{109}$Cd source situated at a calibration point.

\medskip
A modified version of the analysis done in CAST \cite{Garza:2013jg} has been used in the background model of TREX-DM.
This analysis is optimized in discriminating low energy X-ray from external gammas and cosmic muons
and is based on likelihood functions generated by the X-rays' cluster features of a calibration source.
Fixing a total 80\% signal efficiency, the expected background level for an argon-based gas at 10 bar
is {$2 \times 10^2$ keV$^{-1}$ kg$^{-1}$ day$^{-1}$, dominated by the $^{39}$Ar isotope.
If this contribution could be eliminated, the background level will be  $\sim$1  keV$^{-1}$ kg$^{-1}$ day$^{-1}$
and will be limited by the copper vessel and connectors, as shown in figure \ref{fig:Background} (left).
This last value is similar to the one expected in a neon-based mixture. Supossing a 0.4 keVee energy threshold
and a conservative background level of 10$^2$ keV$^{-1}$ kg$^{-1}$ day$^{-1}$,
TREX-DM experiment could be sensitive to a relevant fraction of the low-mass WIMP parameter space
(see figure \ref{fig:Background}, right) including the regions invoked in some interpretations of the hints of the
DAMA/LIBRA and other hints of positive WIMPs signals with an exposure of 1 kg-year. In the near-future, we will study
the z-dependency of cluster features and we will implement a neutron/electron discrimination,
as those made in \cite{Billard:2012jb, Iguaz:2011fi} at low gas pressures.

\begin{figure}[htb!]
\centering
\includegraphics[width=82mm]{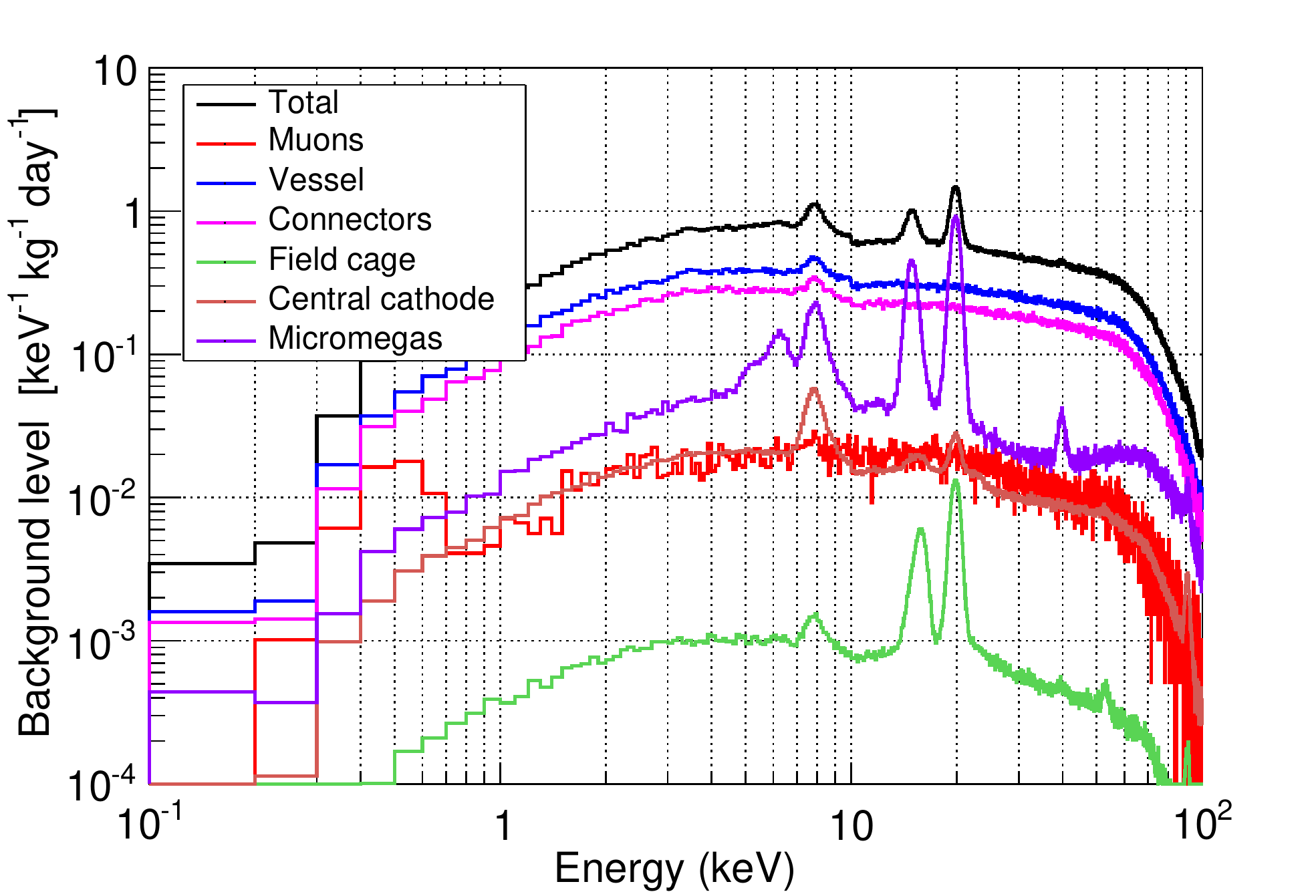}
\includegraphics[width=68mm]{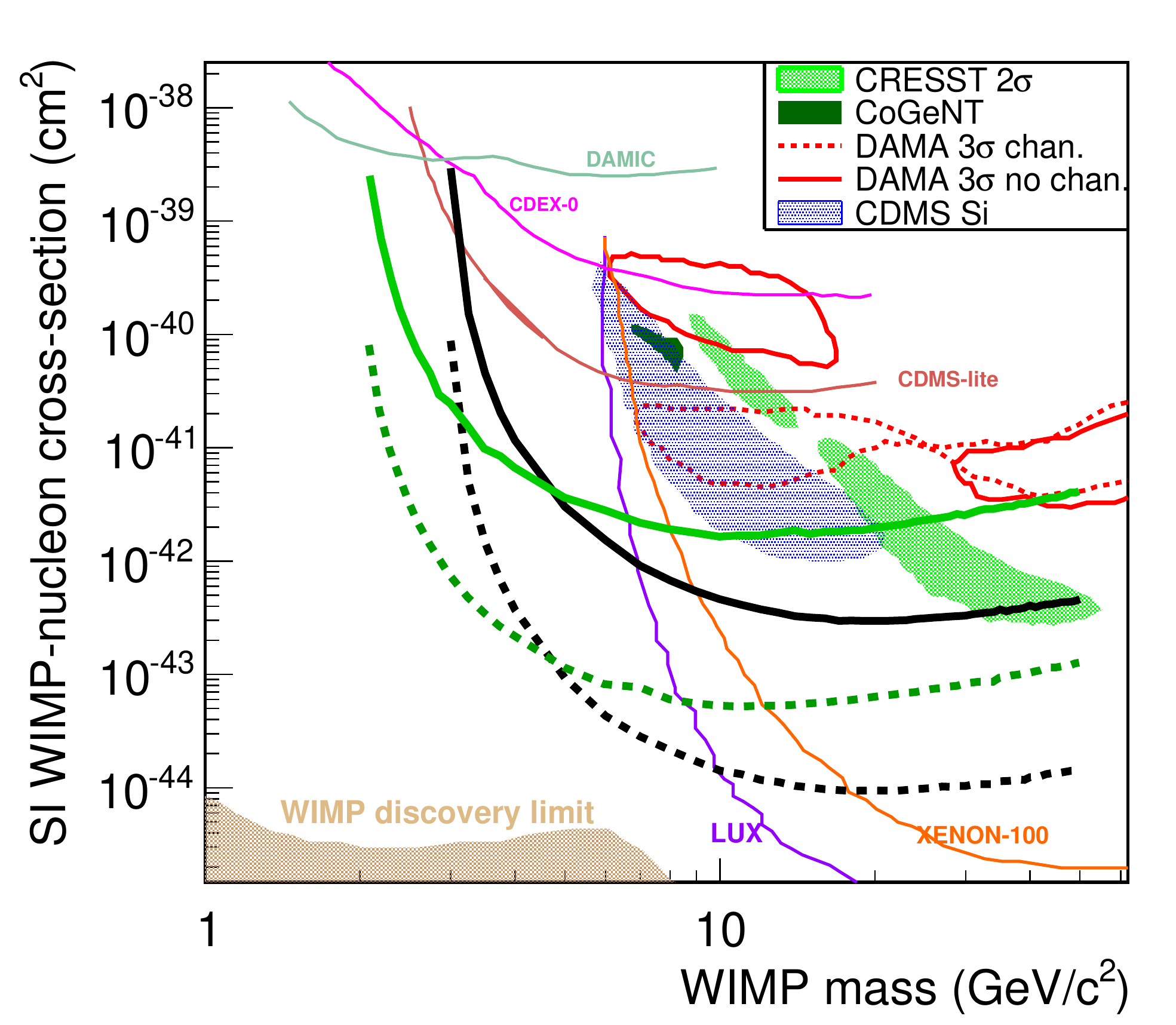}
\caption{Left: Background spectrum expected in TREX-DM experiment (black line) if operated in Ar+2\%iC$_4$H$_{10}$ at 10 bar
in absence of any $^{39}$Ar isotope and installed at LSC.
The contribution of the different simulated components is also plotted: external muon flux (red line),
vessel contamination (blue line), connectors (magenta line), field cage (green line), central cathode (brown line)
and Micromegas detector (purple line) have also been included.
Right: WIMP parameter space focused on the low-mass range. Filled regions represent the values that may explain
the hints of positive signals observed in CoGeNT, CDMS-Si, CRESST and DAMA/LIBRA experiments. The thick lines are
preliminary sensitivity of TREX-DM supossing a 0.4~keVee energy threshold and two different hypothesis
on background and exposure: 100 (solid) and 1(dashed) keV$^{-1}$ kg$^{-1}$ day$^{-1}$,
and 1 and 10 kg-year respectively, and for both argon- (black) and neon-based mixtures (green).}
\label{fig:Background}
\end{figure}

\section{Summary and prospects}
The up-to-now strategy of Dark Matter experiments has been focused on accumulating large target masses
of heavy nuclei target and reducing the background level by a selection of all components and powerful discrimination
methods. However, this approach is not adequate for low-mass WIMPs, as light elements and a low energy threshold are required.
In this context, we present the TREX-DM experiment, a low background Micromegas-based TPC for low-mass WIMP detection.
Its main goal is the operation of an active detection mass $\sim$ 300 g, with an energy threshold of 0.4~keVee or below
and fully built with previously selected radiopure materials.

\medskip
The experiment consists of a copper vessel divided into two active volumes, each of them equipped with a field cage
and a bulk Micromegas $25 \times 25$ cm$^2$ bulk detectors. Signals are extracted from the vessel by flat cables
and are read by an AFTER-chip based electronics. Each side can be calibrated at four different points by a $^{109}$Cd source
and the setup is also equipped with other auxiliar systems (gas, pumping, slow-control).
The actual setup is being comissioned and the Micromegas detectors have been characterized in Ar+2\%iC$_4$H$_{10}$ at 1.2~bar.
Both detectors showed a large plateau at the electron transmission curve, compatible with other bulk Micromegas detectors.
However, the maximum gain reached before the spark limit ($10^3$)
and the energy resolution at 22.1~keV ($\sim$20\% FWHM) are modest. We have attributed
these results to a low quantity of quencher in gas and the presence of a high frequency noise.
Nevertheless, the energy threshold was situated at 1~keV.

\medskip
In the near-term, some improvements will be made in the setup like a better detector's grounding
or a filter for the cathode voltage to remove its important contribution to noise observed at higher voltages.
After these changes, Micromegas detectors will be characterized in Ar+2\%iC$_4$H$_{10}$ up to 10 bar and in other light
gases like neon and helium, studying both the energy threshold and the features of low energy X-ray clusters.
In parallel, we are developing a large bulk Micromegas detector fully built in radiopure
materials. This detector could be also read by AGET electronics \cite{Anvar:2011sa},
which may further reduce the energy threshold to values near 100 eV.
These improvements will be commissioned during 2015,
so as the detector may be installed at the LSC during 2016 for a possible physics run.

\medskip
TREX-DM experiment could be sensitive to a relevant fraction of the low-mass WIMP parameter space
(see figure \ref{fig:Background}, right) including the regions invoked in some interpretations of the hints of the
DAMA/LIBRA and other hints of positive WIMPs signals,
for a 0.4~keVee energy threshold and a background level of $\sim10^2$ keV$^{-1}$ kg$^{-1}$ day$^{-1}$,
as obtained in a first background model. This result is based on the Geant4 simulation of the radioactivity of the
detector components, the signal response
of a Micromegas-based TPC and a modified version of CAST analysis, used to discriminate low energy X-rays from muons.
A lower background level is expected with a neutron/electron discrimination method, even if it will be less effective
at high pressures.

\medskip
We dedicate this work to the memory of our dear IRFU/SEDI colleagues Marc Anfreville and Michel Boyer,
who passed away during 2014.

\ack
We acknowledge the Micromegas workshop of IRFU/SEDI for bulking our detectors and the
Servicio General de Apoyo a la Investigaci\'on-SAI of the University of Zaragoza for the fabrication
of many mechanical components.
We acknowledge support from the European Commission under the European Research Council
T-REX Starting Grant ref. ERC-2009-StG-240054 of the IDEAS program of the 7th EU Framework Program
and the Spanish Ministry of Economy and Competitiveness (MINECO) under grants FPA2011-24058 and FPA2013-41085-P.
F.I. acknowledges the support from the \emph{Juan de la Cierva} program of the MINECO.


\section*{References}
\begin{thebibliography}{99}
\bibitem{Baudis:2012lb}
Baudis L
2012
{\it Physics of the Dark Universe} {\bf 1} 94-108.
\bibitem{Lee:1977bl}
Lee B W and Weinberg S
1977
{\it Phys. Rev. Lett.} {\bf 39} 165.
\bibitem{Baker:2013kb}
Baker K {\it et al.}
2013
{\it Annalen Phys.} {\bf 525} A93-A99.
\bibitem{Bernabei:2013rb}
Bernabei R {\it et al.}
2013
{\it Euro. Phys. J. C} {\bf 73} 2648.
\bibitem{Aalseth:2011cea}
Aalseth C E {\it et al.}
2011
{\it Phys. Rev. Lett.} {\bf 107} (2011) 141301.
\bibitem{Agnese:2013ra2}
Agnese R {\it et al.}
{\it Phys Rev. Lett} {\bf 211} (2013) 251301.
\bibitem{Davis:2014jhd}
Davis J H, McCabe C and Boehm C,
{\it JCAP} {\bf 1408} (2014) 014.
\bibitem{Agnese:2014ra}
Agnese R {\it et al.}
2014 
arXiv:1410.1003.
\bibitem{Akerib:2014dsa}
Akerib D S {\it et al.}
2014
{\it Phys. Rev. Lett.} {\bf 112} 091303.
\bibitem{Agnese:2013ra}
Agnese R {\it et al.}
2014
{\it Phys. Rev. Lett.} {\bf 112} 041302.
\bibitem{Daw:2012ed}
Daw E {\it et al.}
2012
{\it Astr. Phys.} {\bf 35} 397.
\bibitem{Santos:2012ds}
Santos D {\it et al.}
2012
{\it EAS Publication Series} {\bf 53} 25.
\bibitem{Ahlen:2013sa}
Ahlen S {\it et al.}
2010
{\it Int. Jour. Mod. Phys. A} {\bf 25} 1.
\bibitem{Billard:2012jb}
Billard J, Mayet F and Santos D,
2012
{\it JCAP} {\bf 07} 020.
\bibitem{Billard:2012jb2}
Billard J, Mayet F and Santos D
2012
{\it JCAP} {\bf 04} 006.
\bibitem{Giomataris:1995fq} 
Giomataris Y {\it et al.}
1996
{\it Nucl.\ Instrum.\ Meth.\  A} {\bf 376} 29.
\bibitem{Giomataris:2006yg}
Giomataris I {\it et al.}
2006
{\it Nucl. Instrum. Meth. A} {\bf 560} 405.
\bibitem{Andriamonje:2010sa}
Andriamonje S {\it et al.}
2010
{\it JINST} {\bf 5} P02001.
\bibitem{Cebrian:2011sc}
Cebri\'an S {\it et al.}
2011
{\it Astropart. Phys.} {\bf 34} 354.
\bibitem{Aznar::2013fa}
Aznar F {\it et al.}
2013
{\it JINST} {\bf 8} C11012.
\bibitem{Aune:2014sa}
Aune S {\it et al.}
2014
{\it JINST} {\bf 9} P01001.
\bibitem{Alvarez:2014va}
Alvarez V {\it et al.}
2014
{\it JINST} {\bf 9} P03010.
\bibitem{Gerbier:2014gg}
Gerbier G {\it et al.}
2014
arXiv:1401.7902.
\bibitem{Iguaz:2011fa}
Iguaz F J {\it et al.}
2011
{\it JINST} {\bf 6} P07002.
\bibitem{Baron:2008pb}
Baron P {\it et al.}
2008
{\it IEEE Trans. Nucl. Sci.} {\bf 55} 1744.
\bibitem{Baron:2010pb}
Baron P {\it et al.}
2010
{\it IEEE Trans. Nucl. Sci.} {\bf 57} 406.
\bibitem{TREX}
http://gifna.unizar.es/trex/.
\bibitem{Peiro:2013ap}
Peir\'o A
2013
University of Zaragoza
{\it Preprint:} http://zaguan.unizar.es/record/12149.
\bibitem{Iguaz:2012fi}
Iguaz F J, Ferrer-Ribas E, Giganon A and Giomataris I
2013
{\it JINST} {\bf 7} P04007.
\bibitem{Cebrian:2013sc}
Cebrian S {\it et al.}
2013
{\it JINST} {\bf 8} P01012.
\bibitem{Acosta:2008da}
Acosta-Kane D {\it et al.}
2008
{\it Nucl.\ Instrum.\ Meth.\  A} {\bf 587} 46.
\bibitem{Pokratenko:2000op}
Pokratenko O A {\it et al.}
2000
{\it Phys. At. Nucl.} {\bf 63} 1282.
\bibitem{Agostinelli:2003sa}
Agostinelli S {\it et al.}
2003
{\it Nucl.\ Instrum.\ Meth.\  A} {\bf 506} 250.
\bibitem{Tomas:2013at}
Tomas A
2013
University of Zaragoza
{\it JINST} {\bf TH} 001.
\bibitem{Garza:2013jg}
Garza J G {\it et al.}
2013
{\it JINST} {\bf 8} C12042.
\bibitem{Iguaz:2011fi}
Iguaz F J {\it et al.}
2012
{\it Phys. Proc.} {\bf 37} 1079.
\bibitem{Anvar:2011sa}
Anvar S {\it et al.}
2011
{\it IEEE NSS/MIC} 745.
\end{thebibliography}
\end{document}